# Computing the band structure and energy gap of penta-graphene by using DFT and $G_0W_0$ approximations

H. Einollahzadeh[1], R. S. Dariani[1,*], and S. M. Fazeli[2]

[1]Department of Physics, Alzahra University, Tehran, 1993893973, Iran

[2]Department of Physics, University of Qom, Qom, 3716146611, Iran

*Corresponding author. Tel: +98 21 85692646; fax: +98 21 88613935

Email address: dariani@alzahra.ac.ir

**Abstract**

In this paper, we consider the optimum coordinate of the penta-graphene. Penta-graphene is a new stable carbon allotrope which is stronger than graphene. Here, we compare the band gap of penta-graphene with various density functional theory (DFT) methods. We plotthe band structure of penta-graphene which calculated with the generalized gradient approximation functional HTCH407, about Fermi energy. Then, one–shot GW ($G_0W_0$) correction for precise computations of band structure is applied. Quasi-direct band gap of penta-graphene is obtained around 4.1-4.3eV by $G_0W_0$ correction. Penta-graphene is a semiconductor and can be expected to have broad applications in future, especially in nanoelectronics and nanomechanics.

**Introduction**

Carbon is one of the most compatible elements in the environment.Carbon has various allotropes including graphite, diamond, carbon nanotube, fullereneand graphene. Among these structures, two-dimensional carbon structures, havebeen attracted special interest, because of the vast applications of two-dimensional systems [1-2].

In 2004, graphene created via micromechanical cleavage of graphite,by a group in Manchester[3].Graphene is a flat surface of carbon atoms and two- dimensional systems which has three dimensional sp$^2$ hybridized carbon. This nanostructure has widespread applications in nanotechnology for its particular structure and specific electronic properties. Graphene has two atoms in its unit cell which are located in honeycomb structure. It has a linear-relativistic-like band structure consisting of two cones that meet at the so-called "Dirac-point". Energy gap of graphene is zero, which leads to semi-metal behavior in graphene [4]. Graphene has significantly different properties. It conducts electricity better than gold or silver and heat better than diamond. Graphene is transparent and one of the most flexible material. It is good device for supercapacitors. Therefore, graphene could revolutionize industries, because of its unique properties [5]. The successful production of graphene has encouraged scientists to investigate novel graphene allotropes [6-8].

Pentagon and hexagon are two basic structures for carbon nanostructure. The topological defect or geometrical frustration in carbon, are usually has a pentagon form [9-10].

Recently, considerable efforts focused on stabilization of pentagon-based carbon. In early 2015, researchers at Virginia, China and Japan have proved that a new two- dimensional carbon allotrope which merely consists of pentagon structures could exist theoretically [11]. They showed that this structure has got dynamical, thermal and mechanical stability. Penta-graphene is extracted from T12-carbon phase [12]. This phase is obtained by heating an interlocking-hexagon-based metastable carbon phase at high temperature [11].

In present work, the band structure of penta-graphene is computed by using different density functional theory (DFT) approximations. We also report indirect band gap by different DFT functional methods and quasi-direct band gap of penta-graphene is showed. As we know, DFT band gap value is weak in semiconductors. Hence, we use $G_0W_0$ for improving the band gap value.

**Computational method**

In this computation, we use ABINIT code, this code is based on pseudopotentials and plane waves. By using ABINIT code, we can compute total energy, charge density and electronic structure. Exited states can be measured from time dependent density functional theory and many body perturbation theories [13].

DFT is the most successful approach which is broadly used in condensed matter and computational physics for describing properties of condensed matter systems. The main idea of DFT is to describe a many body interacting system through the particle density [14]. DFT calculations use Kohn-Sham (KS) method. The Kohn-Sham applies standard independent-particle methods for estimating the exact density and energy of a many-body electron problem [15]. An accurate DFT computation is possible via approximation exchange-correlation (xc) potential. Although calculatingof the exact xc potential is impossible, there are some approximations for estimating it; One of them is local density approximation (LDA). In LDA, the xc energy per electron at point **r** substitutes by a homogenous gas which has the same electron density at point **r**.Another approximation is generalized gradient approximation(GGA), which uses both $n(r)$ and$|\nabla n(r)|$. The LDA and theGGA functional underestimate Kohn-Shamband gaps systematically [16].

DFT is concerning only ground state properties. Computing band gap needs at least an electron in the conduction band, which is an exited state and therefore is not within ground state. Thus, it is necessary to go beyondDFT.

GW approximation is a many body perturbation theory and is based on Green function, quasiparticle conception and an expansion of exchange correlation self-energy. It is usually carried within RPA[1]which in this approximation the vertex function in reciprocal space is estimated by the screened coulomb interaction [17]. In the GW approximation, the self-energy operator is given by the product of the one-electron Green function (G) and the screened coulomb interaction ($\Sigma=iGW$). Usually, the GW calculation is performed using perturbation theory from LDA (GGA) result, which provides a quite efficient starting point.

$$\varepsilon_{nk}^{GW} = \varepsilon_{nk}^{KS} + Z\langle nk|\Sigma^{GW}\varepsilon_{nk}^{KS} - v_{xc}^{LDA}|nk\rangle \quad (1)$$

The band gap result in the GW quasiparticle energies are in excellent agreement with the available experimental data [18].

Computing the full self-consistent GW method for real systems is very controversial, so that the most usual approach is using the best approximation for G and W as a starting point (the so-called one-shot GW or $G_0W_0$ )[19]. $G_0W_0$ is a first order correction to a single particle Hamiltonian. Here, we use one-shot GW ($G_0W_0$) for improving band structure.

---

[1]Random Phase Approximation

**Results and discussion**

The optimum crystal structure of two-dimensional penta-graphene is prepared by exfoliating from T12-carbon. This structure has symmetry p-42$_1$m and it can be described by tetragonal lattice, with six carbon atoms per unit cell where a=b=3.64 Å are the optimized lattice constants. The thickness of this two-dimensional sheet is 1.2Å. This structure has two hybridization sp$^2$ and sp$^3$. The sp$^3$ and sp$^2$ hybridizations have nominated by C1 and C2, respectively, that C1: C2 ratio is 1:2. The band length of single bond (C1-C2) is 1.55Å and the band length of double bond (C2-C2) is 1.34Å.

By applying these specifications, we find optimal structure with ABINIT code. Our optimized lattice is figured out in Fig.1. An optimized lattice constant is obtained 3.64Å which agrees with previous calculations [11].

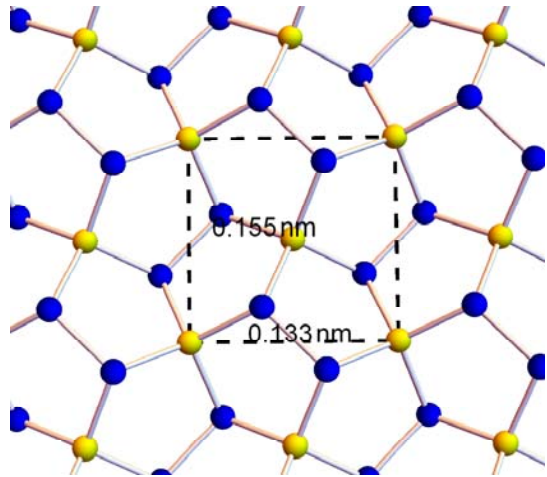

Fig. 1. Top view of optimized structure of penta-graphene, yellow, blue sphere represent C1 and C2; respectively.

In the following, we use DFT-GGA and DFT-LDA calculation in the form of different exchange-correlation (xc) functions (based on the Kohn–Sham eigenvalues). In this computation, vacuum alone the *z* direction is 10 Å. This parameter isconverged, so that the penta-graphene layer becomes isolated. The plane wave basis set is50 Ha (Hartree) that total energy converges to 1meV/atom. We use Martins-Trouiller pseudopotential [20]. The k point set is based on a 18 18 1 Monkhorst pack grid [21]. The band structure of penta-graphene is determined by DFT-LDA approximation, therefore we could compute band gap of penta-graphene in this approximation. Our result is showed in table 1.Band gap in this computation is around 2.3 eV (2.21-2.36 eV).

Table 1. The indirect energy gap of penta-graphene by diffferent DFT funtionals.

| ABINIT code | $E_g$ | Functional |
|---|---|---|
| Ixc=1 | 2.22 eV | LDA-Teter Parametrization |
| Ixc=5 | 2.21 eV | LDA-Hehin-lundqvist |
| Ixc=15 | 2.29 eV | GGA-RPBE |
| Ixc=23 | 2.25 eV | GGA Z.Wu |
| Ixc=24 | 2.22 eV | GGA C09x |
| Ixc=26 | 2.34 eV | GGA,HTCH147 |
| Ixc=27 | 2.36eV | GGA,HTCH407 |

As we said, the penta-graphene has a tetragonal lattice; the high symmetry points in tetragonal are ($\Gamma$=(0,0,0), X=(0,1/2,0) and M=(1/2,1/2,0)), therefore we have plotted the band structure of penta-graphene along the symmetry line $\Gamma$-X-M-$\Gamma$. For instance, we have plotted the band structure for one of the functional of table 1, HTCH407 GGA in Fig.2(a).

In Fig. 2, we have showed that the penta-graphene is a semiconductor with a indirect band gap, since the maximum of valance band (VBM) is in $\Gamma$-X path and the minimum of conduction band lies on M-$\Gamma$ path.

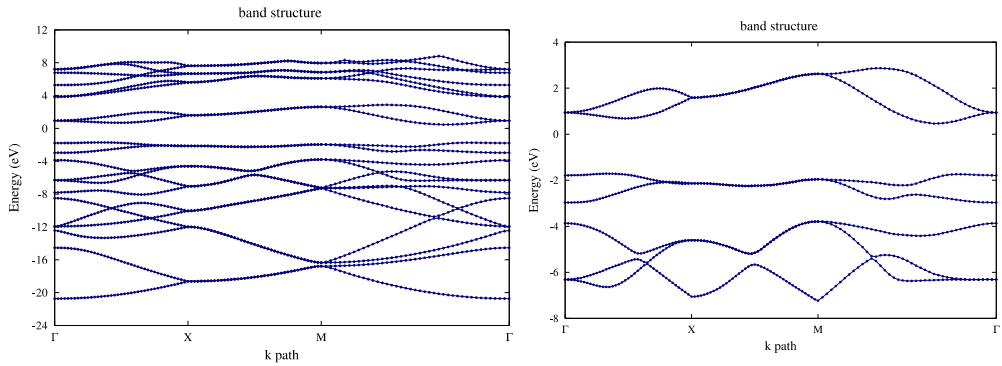

Fig. 2. (a) The band structure of penta-graphene calculated by using HTCH407 GGA for 20 band. (b)theband structure of penta-graphene around Fermi energy

We have used Martins-Trouiller pseudopotentials in our calculations, by using these pseudopotentials [22], two electrons of 2s are applied in potential, and so by omitting these electrons, we have four electrons per carbon atoms. Penta-graphene has six carbons in its unit cell; so that, we have 24 electrons in unit cell and accordingly penta-graphene has 12 filled bands. Therefore, energy gap is located between the twelfth and the thirteenth levels. For better specifying the energy gap, we have plotted the band structure around Fermi energy in Fig.2 (b).

It is obvious in Fig. 2 that sub-VBM on the M-Γ path is very close to the true VBM in energy. Therefore, penta-graphene can also be investigated as a quasi-direct band gap semiconductor. The difference between quasi-direct band gap and indirect band gap is about 0.01 eV.

We knowDFT is not accurate in predicting band gap of semiconductors, so we used $G_0W_0$ approximation for obtaining a better value of energy gap. In this approximation, at first step the ground state energy and electronic density are calculated. Then, the Kohn-Shamelectronic structure is used for $G_0W_0$ excited calculation. This means that the Green function G and dynamically screened interaction W are made by electronic structure of KS.

Here, the plane-wave basis set has been expanded to an energy cut-off 20 Ha (That converges the total energy to 1meV/atom) and calculations converge at n=50 for computing W and self-energy. The k-point set is based on a 18 18 1 Monkhorst pack grid. We compute the band structure of penta-graphene, for three different functional (table 2). The effect of $G_0W_0$ correction is opening the band gap about 2 eV.

Table 2. The quasi-direct band gap of penta-graphene.

| ABINIT code | Functional | Quasi-direct band gap(GW) |
| --- | --- | --- |
| Ixc=1 | LDA-Teter Parametrization | 4.10eV |
| Ixc=15 | GGA-RPBE | 4.14 eV |
| Ixc=27 | GGA,HTCH407 | 4.28eV |

We use Mathematica interpolation for plotting full band structure. We have compared, for instance, the DFT and the $G_0W_0$ approximation in GGa-RPBE functional of penta-graphene in Fig .3

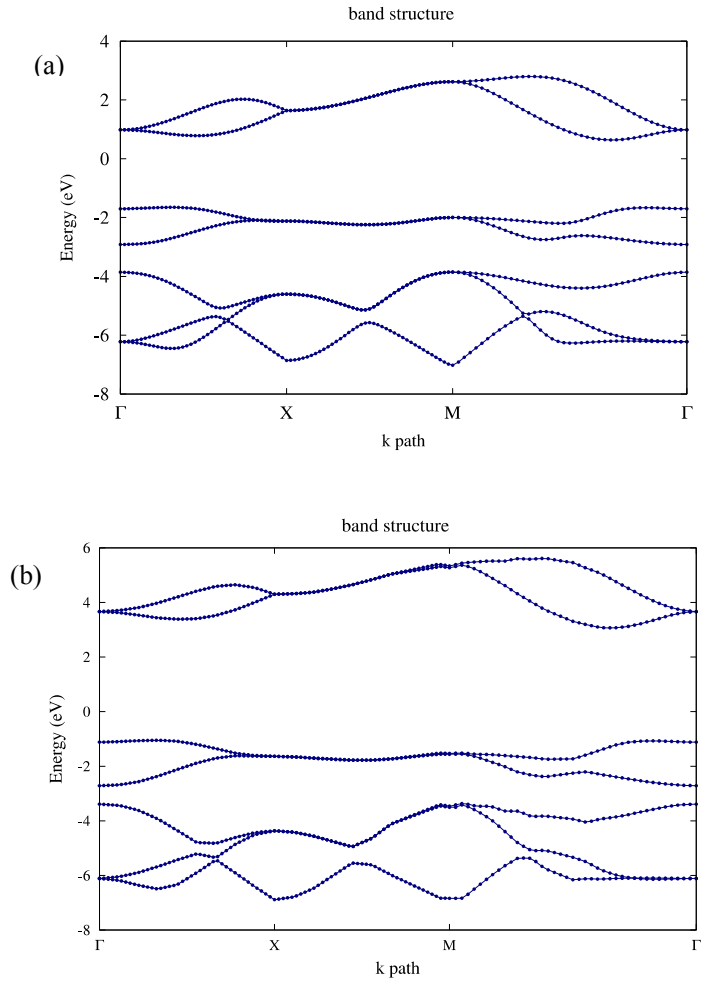

Fig. 3. The band structure of penta-graphene by using GGa-RPBE functional: (a) The DFT (b) The $G_0W_0$ approximation.

From Fig. 3. We see that the penta-graphene is a new carbon allotrope, with indirect band gap or quasi direct band gap about 4.1-4.3 eV.

**Conclusions**

In this paper, the band structure of penta-graphene has been investigated. This new structure, which is made by theoretical model, has got many advantages in comparison to graphene. It is not only dynamically and mechanically stable, but also can tolerate up to 1000K temperature. At first by using the ABINIT code, the optimum structure and the optimum lattice constant has been acquired. This structure has two types of carbon with $sp^2$ and $sp^3$ hybridizations. Then, by applying DFT with various densities functional, the indirect band gap has been obtained. Since the sub-VBM on the M-Γ path is very close to the true VBM in energy, the quasi-direct band gap could be related to penta-graphene. The $G_0W_0$ approximation has been applied for finding accurate band gap. Finally, the band gap has been obtained about 4.1-4.3 eV. Consequently, we conclude that the penta-graphene is a semiconductor with nearly big band gap. From this band gap, it can be expected to have interesting applications in optoelectronics.

**References**

[1] A. Hirsch, Nature material 11 (2010) 868-871.

[2] W. Yihong, Sh.Zexiang,and Y.Ting, "Two-Dimensional carbon: Fundamental Properties,Characterization and Applications",CRC Press(2014).

[3] K. S. Novoselov, A. K. Geim, S. V. Morozov, D. Jiang, Y. Zhang, S. V. Dubonos, I. V. Gregorieva, and A. A. Firsov,Science306 (2004) 666.

[4] M. I. Kasnelson, "Graphene: Carbon in two dimensions", Cambridge University Press(2012).

[5] A. F. Morpurgo, Nature Physics 11(8) (2015) 625-626.

[6] Y. Li, L. Xu, and H. Liu, Chem. Soc. Rev. 43(8) (2014) 2572-2586.

[7] A. N. Enyashin and A. L. Ivanovskii, Physica Status Solidi (b) 248(8) (2011) 1879-1883.

[8] L. C. Xu, R. Z. Wang, M. S. Miao, X. L. Wei, Y. P. Chen, H. Yan, W. M. Lau, L.M. Liu, and Y. M. Ma, Nanoscale 6(2) (2014) 1113-1118.

[9] M. Deza, P. Fowler, M. Shtogrin, and L. Vietze, J. Chem.Inf.Comput. Sci. 40(6) (2000) 1325-1332.

[10] F. B. Machado, A. J. Aquino, and H. Lischka, Physical Chemistry Chemical Physics 17(19) (2015) 12778-12785.

[11] S. Zhang, J. Zhou, Q. Wang, X. Chen, Y. Kawazoe, and P. Jena, Proceedings of the National Academy of Sciences 112(8) (2015) 2372-2377.

[12] S. Zhang, Q. Chen, X. Chen and P. Jena, , Proceedings of the National Academy of Sciences, 110(47) (2013) 18809-18813.

[13] X. Gonze, B. Amadon, P. M. Anglade, J. M. Beuken, F. Bottin, P. Boulanger, F. Bruneval, D. Caliste, R. Caracas, M. Cote, T. Deutsch, L. Genovese, Ph.Ghosez, M. Giantomassi, S. Goedecker, D. R. Hamann, P. Hermet, F. Jollet, G. Jomard, S. Leroux, M. Mancini, S. Mazevet, M. J. T. Oliveira, G. Onida, Y. Pouillon, T. Rangel, G. M. Rignanese, D. Sangalli, R. Shaltaf, M. Torrent, M. J. Verstraete, G. Zerah, and J. W. Zwanziger, Comput. Phys. Comm. 180(12) (2009) 2582-2615.

[14] M. P. Marder, "Condensed Matter Physics", the University of Texas at Austin (2000).

[15] R. M. Martin, "Electronic Structure: Basic Theory and Practical Methods", Cambridge University Press (2004).

[16] C. Delerue and M. Lannoo, "Nanostructures Theory and Modeling", Springer (2004).

[17] W. G. Aulbur, L. Jonsson and J. W. Wilkins," Quasiparticles calculations in solids" Columbus (2000).

[18] S. Korbel, P. Boulanger, L. Duchemin, X. Blase, M.A. Marques, S. Botti, Journal of Chemical Theory and Computation 10(9) (2014) 3934-3943.

[19] R. Laaner, Journal of physics: Condensed Matter 26(12) (2014) 125503.

[20] N. Troullier, J. L. Martin, Phys. Rev. B 43(3) (1991) 1993-2006.

[21] H. J. Monkhorst and J. D. Pack, Phys. Rev. B 13(12) (1976) 5188-5192

[22] J. Kohanoff, "Elaectronic structure calculations for solids and molecules: theory and computational methods", Cambridge University Press (2006).